\documentclass[12pt,preprint]{aastex}

\begin{document}

\title{Solid Quark Stars?}

\author{R. X. Xu}
\affil{School of Physics, Peking University, Beijing 100871,
China}

\begin{abstract}

It is conjectured that cold quark matter with much high baryon
density could be in a solid state, and strange stars with low
temperatures should thus be solid stars.
%
%
The speculation could be close to the truth if no peculiar
polarization of thermal X-ray emission (in, e.g., RXJ1856), or no
gravitational wave in post-glitch phases, is detected in future
advanced facilities, or if spin frequencies beyond the critical
ones limited by $r-$mode instability are discovered.
The shear modulus of solid quark matter could be $\sim 10^{32}$
erg/cm$^3$ if the kHz QPOs observed are relevant to the
eigenvalues of the center star oscillations.

\end{abstract}


\section{Introduction}

The gauge theory of strong interaction, the quantum chromodynamics
(QCD), is still developing, which is nevertheless well known to
have two general properties: asymptotic freedom in smaller scales
($\sim 0.1$ fm) and color confinement in larger scales ($\sim 1$
fm).
These result in two distinct phases depicted in the diagram in
terms of temperature $T$ vs. baryon chemical potential $\mu_{\rm
b}$: the hadron gas (in the region of low $T$ {\em and} $\mu_{\rm
b}$) and the quark matter (of high $T$ {\em or} $\mu_{\rm b}$).
The former has been studied well in nuclear physics, but the later
is expected to be a new phase which has never been found with
certainty yet.
There are actually two kinds of quark matter in research: the
temperature-dominated phase which may appear in the early universe
or in relativistic heavy-ion colliders, and the density-dominated
phase, speculated only in astrophysical compact objects, which is
the focus of this letter.

One can not naively think that the density-dominated quark matter
is simply a Fermi-gas with weak interaction.
In fact, the attraction, no matter how weak it is, between quarks
renders the Fermi-sea unstable to the quark Cooper pairing of the
BCS type in a color superconducting state (Alford, Bowers \&
Rajagopal 2001, and references therein).
This kind of condensation in momentum space takes place in case of
same Fermi momenta; whereas ``LOFF''-like state may occur if the
Fermi momenta of two (or more) species of quarks are different.
For 3 flavors of massless quarks, all nine quarks may pair in a
pattern which locks color and flavor symmetries, as called
color-flavor locked (CFL) state.

We suggest, however, an alternative ``condensation'' in {\em
position} space for quark matter with high baryon density, and
discuss the possibility of such matter in a solid state and its
various astrophysical implications. We are concerned about the
most probable compact objects composted entirely of deconfined
light quarks, the so called strange stars (e.g., Glendenning
2000).
In fact a few authors had got involved in this possibility more or
less.
By extending the bag-model study on any color-singlet multi-quark
system, Jaffe (1977) predicted an S-wave flavor-singlet state of
6-quark cluster ($H$ particle).
An 18-quark cluster (called quark-alpha, $Q_\alpha$) being
completely symmetric in spin, color, and flavor space was also
proposed (Michel 1988), which could be essential to reproduce
pulsar glitches by modifying significantly the strange star
structures (Benvenuto, Horvath \& Vucetich 1990).

\section{Can quark matter be solid?}

There are virtually two ingredients which affect the formation of
quark cluster in quark-gluon plasma: the Pauli's exclusion
principle and the interactions.
Suppose we ``turn-off'' the interactions at first.
Let's consider two quarks for simplicity. The wave function of
quarks is $\Psi(1,2)=\Psi^{\rm inner}(1,2)\Psi^{\rm
position}(1,2)$, where $\Psi^{\rm inner}(1,2)$ describes particles
in inner space while $\Psi^{\rm position}(1,2)$ in position space.
Due to the identical principle of fermions, these two quarks
exchange asymmetrically, i.e., $ P_{12}\Psi(1,2)=-\Psi(1,2)$, with
$P_{12}=P^r_{12}P^\sigma_{12}P^f_{12}P^c_{12}$, where
$r,\sigma,f,c$ denote the position, spin, flavor, and color
freedoms, respectively.
Quarks have a tendency to condensate in position space if
$\Psi^{\rm position}(1,2)$ is exchange-symmetric,
\begin{equation}
\Psi^{\rm position}_{\rm AB}(1,2)={1\over\sqrt{2}}
[\Phi_{\rm A}(1)\Phi_{\rm B}(2)+\Phi_{\rm B}(1)\Phi_{\rm A}(2)], %
\label{Psi}
\end{equation}
while $P_{12}\Psi^{\rm inner}(1,2)=-\Psi^{\rm inner}(1,2)$.
Eq.(\ref{Psi}) shows that the possibility of one quark
``hobnobbing'' with another is high. But this possibility is
almost zero if $\Psi^{\rm position}(1,2)$ is
exchange-antisymmetric.
In order to see the significance of this exchange effect, Zhang \&
Yu (2002) have calculated the matrix of operator $P_{12}$, and
find that, for a system with two baryons of S-wave, it is favored
to form some bound states of 6-quark cluster, e.g.,
($\Omega\Omega$), ($\Delta\Delta$), etc.

Whether quark clusters can form relies also on the model-dependent
color, electro-magnetic, and weak interactions, particularly the
dominate color one.
We therefore conjecture the existence of $n$-quark clusters in
quark matter, leaving $n$ as a free parameter, possibly from 1 to
18, or to be even larger.
Note that $n$ may depend on temperature and baryon density, and
that more than one kind of quark clusters may appear at a certain
phase (i.e., quark clusters with different $n$'s could coexist).
Although these hypothetic quark clusters may not form in
terrestrial laboratory environments with low baryon density (or
$\mu_{\rm b}$), it is very likely that quarks could be clustered
in the ``deconfined'' phase, especially in the strange quark
matter assumed to be absolutely stable.

The next step is to investigate the melting temperature $T_{\rm
m}^n$ for the component of $n$-quark clusters. The quark matter
could be solid as long as the temperature $T$ is less than one of
$T_{\rm m}^n$'s.
The $n$-quark clusters might be divided into two classes: color
singlet or not. The former would be the quark analog of molecules
(e.g., H$_2$O and CO$_2$), the short-distanced color force (the
color ``Van der Waals force'') between which could crystallize the
clusters at low $T$ (as an experiment example of H$_2$O, water
boils at 100$^{\rm o}$C when density $\sim$1 g/cm$^3$, but freezes
even at 300$^{\rm o}$C when density $\sim$1.5 g/cm$^3$), and the
later is of ionic lattice (e.g., NaCl, ZnS, something like plasma
{\em solid} rather than fluid) when $T<T_{\rm m}^n$. Due to the
color screen in quark-gluon plasma, $n$-quark clusters with color
charge can not interact in long range but is also short-distanced.
Phase transition, which could be more than once, occurs when one
component of the $n$-quark clusters begins to be fixed at certain
lattices as the matter cools.
We may assume a Van der Waals-type potential between $n$-quark
clusters since both kinds of clusters interact in short distance.
The melting temperature $T_{\rm m}\sim \varepsilon_{\rm B}$, where
$\varepsilon_{\rm B}$ is the Van der Waals binding energy.
For conventional solid matter bound with the electric interaction
(Weisskopf 1985), $\varepsilon_{\rm B}\sim (1-10)$ kcal/mol [or
$\sim (0.05-0.5)$ eV], and the distance between two nearby atoms
is $r_{\rm a}\sim (2-4)$ \AA. For two unit charges at this
separation, the interaction energy could be $\epsilon\sim
e^2/r_{\rm a}\sim (7-4)$ eV. Therefore, due to the electric screen
effect, the binding reduces by a factor of $S_{\rm
e}=\epsilon/\varepsilon_{\rm B}\sim (10-200)$.
If the interaction between quarks could be described as a
Coulomb-like potential (e.g., Lucha et al. 1991), $V(r)\sim
\alpha_{\rm s}/r$ with $\alpha_{\rm s}\equiv g_{\rm s}^2/4\pi$ and
$g_{\rm s}$ the strong coupling constant, the Van der Waals-type
binding energy between $n$-quark clusters could be
$\varepsilon_{\rm B}\sim \alpha_{\rm s}(n_{\rm b}/n)^{1/3}/S_{\rm
c}$, where $S_{\rm c}$ is the correspond color screen factor,
$n_{\rm b}$ the baryon number density.
For quark matter with about two times of normal nuclear matter
density, $n_{\rm b}\sim 0.3~{\rm fm}^{-3}$, $n\sim 10$,
$\alpha_{\rm s}\sim 1$, and $S_{\rm c}\sim S_{e}$, we have $T_{\rm
m}^n\sim \varepsilon_{\rm B}\sim (0.3-6)$ MeV which is much larger
than the surface temperatures observed ($\sim 100$ eV, e.g., Xu
2002) of potential strange star candidates.
A newborn strange stars could thus be solidified soon after
supernova.
The nature of this kind solid quark matter may depend on the quark
cluster structures (Liu et al. 2003) too.


\section{Tests for the solid strange star ideal}

How can we make sure whether a strange star candidate is solid or
not? Although the possibility of solid cold quark matter can not
be ruled out, unfortunately, due to the complex nonlinearity of
QCD, it is almost impossible to draw a certain conclusion yet from
the first principles.
Nonetheless, possible critical observational tests are addressed,
which may finally present a clear answer to the question.

{\em Polarization of thermal photons?}
For hot bare strange star (BSS) with temperature $T>T_{\rm u}\sim
10^9$ K ($\sim 0.1$ MeV), the mechanism, proposed by Usov (1998),
of pair production and their annihilation works (e.g., Usov 2001).
Such a hot BSS is in a fluid state if $T>T_{\rm m}$.
However, a cold strange star with $T<T_{\rm m}$ is solid, and the
Usov mechanism doesn't work if $T<T_{\rm u}$. The photon
emissivity may not be negligible in this case since the opinion,
that the BSS surfaces should be very poor radiators at $T\ll
\hbar\omega_{\rm p} \sim 20$ MeV ($\omega_{\rm p}$ is the plasma
frequency), is for fluid plasma rather than solid one.

In a BSS, although part (or all) of the quarks are clustered in
fixed lattices, electrons with number density $n_{\rm e}\sim
10^{-4}n_{\rm b}$ are free (the density of free electrons could be
much smaller if part of electrons and clusters form ``atoms'' or
``ions''). The interactions of an electron with another one, or
with the lattices ($n$-quark clusters), could be responsible to
the thermal photon radiation.
Electrons are in the levels of the energy bands (rather than
discrete levels) because of their motion in a periodic lattice.
This may result in a featureless spectrum as observed (Xu 2002).
The thermal emission of such a solid could be analogous to that of
metals (Born \& Wolf 1980) to some extent.
According to Kirchhoff's law of thermal radiation, the spectral
emissivity
\begin{equation}
\psi(\nu, T)=\alpha(\nu, T)\cdot B(\nu, T),
\label{psai}
\end{equation}
where $\alpha(\nu, T)$ is the spectral absorptance, and $B(\nu,
T)$ is the Plank function of blackbody thermal radiation.
Therefore it is essential to calculate $\alpha(\nu, T)$ in order
to have a thermal spectrum of BSSs.
We are embarrassed if we try to obtain an exact absorptance
$\alpha(\nu, T)$ from QCD phenomenological models.
Nevertheless, $\alpha(\nu, T)$ can be calculated in classical
electrodynamics, which is a function of electric conductivity
(Born \& Wolf 1980).

Within the realm of solid bare strange stars, Zhang et al. (2003)
have fitted well the 500ks Chandra LETG/HRC data for RX
J1856.5-3754 in the spectral model of Eq.(\ref{psai}), with a
spectral absorptance of metals (Born \& Wolf 1980),
$ \alpha(\nu)=1-[2\sigma/\nu+1-2\sqrt{\sigma/\nu}]/[2\sigma/\nu
+1+2\sqrt{\sigma/\nu}]$,
where $\sigma$ is the conductivity, which could be a function of
temperature $T$. They found a low limit of electric conductivity
of quark matter $\sigma>1.8\times 10^{17}$ s$^{-1}$, with the
fitted radiation temperature $T_\infty\sim 60$ eV and radius
$R_\infty>5.3$ km.
If the electrons near the Fermi surface are responsible for the
conduction, one has $\sigma=n_{\rm e}e^2\tau/m_*$, where the
electron number density $n_{\rm e}\sim 10^{34}$ cm$^{-3}$, $e$ the
electron charge, $\tau$ the relaxation time, and $m_*$ the
effective electron mass. For the case of free degenerated gas, we
have $\tau>\sim 1.2\times 10^{-21}$ s.
If electron-electron collisions are responsible for the
conduction, one can obtain the relaxation time (e.g., Flowers \&
Itoh 1976; Potekhin et al. 1997).
Using the Appendix A.1 by Potekhin et al. (1997) as an estimate,
one has $\tau\sim 2.3\times 10^{-16}$ s for typical parameters.
The value could be larger or smaller if 1, part of electrons are
captured by lattices by Coulomb force, and 2, other interactions
(e.g., electron-phonon, electron-defect) are included,
respectively.
In summary, the conductivity fitted could be reasonable.

None-atomic feature is found in the thermal spectra detected,
which is suggested to be evidence for bare strange stars (Xu 2002;
note: the absorptions in 1E 1207.4-5209 are very probably
associated with cyclotron lines, see Bignami et al. 2003, Xu et
al. 2003).
How can a neutron star reproduce such a spectrum?
Strong magnetic fields may help, for instance, RXJ1856 to do this,
since a condensation transition in the outermost layers may occur
if RXJ1856 has a high magnetic field ($>10^{13}$G for Fe and
$>10^{14}$G for H atmospheres, respectively; Turolla, Zane \&
Drake 2003).
However, in such a strong magnetic field, the surface thermal
X-ray emission should have a peculiar nature of polarization
(e.g., Gnedin \& Sunyaev 1974; Lai \& Ho 2003: the polarization
plane at $<1$ keV is perpendicular to that at $>3$ keV).
But if RXJ1856 is a solid BSS with normal magnetic field ($\sim
10^{12}$ G), the surface density $\rho>10^{14}$ g/cm$^3$, and the
particle kinematic energy density is thus much greater than the
magnetic one, $\rho c^2\gg B^2$. We expect therefore, for solid
BSSs, that the magnetic effect on the thermal photon emission is
negligible, and that the polarization of X-rays is much {\em
lower}.
These results provide a possibility to test neutron or strange
star models (e.g., for RXJ1856) by X-ray polarimetry in the
future, if the depolarization effect in the magnetosphere is
negligible.
Rapid rotating may smear a possible spectral line, but a
fast-rotating pulsar can hardly be ``dead''.
Further differences between the high-energy emission of a solid
BSS with  $\sim 10^{12}$ G and a neutron star with $\gg 10^{12}$ G
might be possible if vacuum nonlinear electrodynamic effects are
included (Denisov \& Svertilov 2003).


{\em Pulsar timing?}
The spin variation of the earth could be an effective probe to
earth's interior (e.g., the Chandler wobble and the ten-year
timescale change of spin may show various kinds of coupling
between the fluid core and the solid mantle).
It is also possible to investigate pulsar interior by timing, the
great precision of which is unique in astronomy.
Apart from planets, radio pulsars are another example of solid
state matter in the universe for which glitch is clear and direct
evidence.
Radio pulsars, at lest part of them, could be BSSs for the
peculiar nature of drifting subpulses (Xu et al. 1999) though
these phenomena may also be explained if the actual surface
magnetic field at the polar cap of neutron stars is very strong
($\sim 10^{13}$ G) and highly non-dipolar (Gil \& Melikidze 2002).
Furthermore, precessions and glitches observed may be well
understood if pulsars are {\em solid} BSSs.

The equilibrium figures of rotating stars can be approximated by
Maclaurin spheroids. For the earth, such an eccentricity
calculated is $e=0.092$, which is of the same order observed
($e=0.083$).
PSR B1828-11 has a rotation frequency $\omega=2\pi/0.405$
s$^{-1}$, and precesses with probably a frequency $\omega_{\rm
p}=2\pi/500$ day$^{-1}$ (Stairs et al. 2000, Link \& Epstein
2001). In case of homogenous ellipsoid approximation, $\omega_{\rm
p}/\omega\approx 0.5e^2$
(for $e\ll 1$), 
with the eccentricity $e=\sqrt{a^2-c^2}/a$ ($a$, $c$ are the
semimajor and semiminor axes, respectively).
We have $e=1.4\times 10^{-4}$ from the precession observation, but
$e=2.2\times 10^{-3}$ from a calculation of a Maclaurin spheroid
with one solar mass and 10 km in radius.
This discrepancy could be due to: 1, the homogenous approximation,
2, the general relativistic effect, and 3, the strong interaction
if the pulsar is actually a strange star (no eccentricity if the
centrifugal effect is negligible).

The observation of PSR B1828-11 precession means that the vortex
pinning is much weaker than that predicted in pulsar glitch
models.
If PSR B1828-11 is a vortex unpinning neutron star with a
spherical superfluid core, the observation may be of the crust
precession (see Eq.(32) of \S384 in: Lamb 1932).
But if it is with an elliptical fluid core (the crust mass $\sim
10^{-5}M_\odot$ is negligible), the precession frequency could be
(Eq.(42) of \S384 in: Lamb 1932) $\omega_{\rm p}\sim \omega/e\gg
\omega$, which conflicts with observations.
More than these, in both cases, the crust could be deformed
irrecoverably by the inertial force of the fluid core, since the
anelasticity crust have a tendency to spin along one of the
principal inertial axes.

For such a neutron star with radius $R$, the strain $\sim
\dot{\alpha}\tau/\alpha$, and the stress $\sim
(2\pi/3)e^{-2}R^2\rho\alpha^2\omega^2$, with $\alpha$ the angle
between rotational and principal axes and $\tau$ the timescale of
relaxation.
One has thus $\alpha(t)\sim
[(4\pi/3\mu\tau)e^{-2}R^2\rho\omega^2t+\alpha_0^{-1}]^{-1/2}$,
where the shear modulus of neutron star crusts could be (Fuchs
1936)
$\mu^{\rm NS}=2.8\times 10^{28}\delta
Z_{26}^2A_{56}^{-4/3}\rho_{11}^{4/3}~~{\rm erg/cm^3}$
%
($\delta=0.37$, ions have charge $Z=26Z_{26}$ and mass number
$A=56A_{56}$, and the density $\rho=\rho_{11}10^{11}$g/cm$^3$),
and $\alpha_0=\alpha(0)$.
In case of $\alpha_0\sim 3^{\rm o}$ (Link \& Epstein 2001), the
timescale for $\alpha$ change from $\alpha_0$ to $\alpha_0/2$ is
then
\begin{equation}
T_{1/2}^{\rm NS}\sim 9.1\times 10^{-5}\tau^{\rm
NS}\mu_{28}\rho_{11}^{-1},
\label{TNS}
\end{equation}
where $\mu_{28}=\mu/(10^{28}$erg/cm$^3)$.
However for the pulsar being a solid strange star, the
corresponding value is
$T_{1/2}^{\rm SS}\sim 9.1\times 10^3\tau^{\rm
SS}\mu_{32}\rho_{15}^{-1}$,
%
which should be more than 8 orders lager than $T_{1/2}^{\rm NS}$
($\mu^{\rm SS}\sim 10^{32}$ erg/cm$^3$, see Eq.(\ref{muSS}), and
the stress of solid stars $\sim (2\pi/3)R^2\rho\alpha^2\omega^2$).
Such a great difference may be used to test the solid strange star
model by obtaining the beam width change as the precession
amplitude decays, although the absolute timescales of $\tau^{\rm
NS}$ and $\tau^{\rm SS}$ are still uncertain.

Glitches observed may be well understood if pulsars are {\em
solid} strange stars.
As a pulsar slows down and therefore the centrifugal force
decreases, stresses develop due to (1) reducing the stellar
volume, and (2) a less oblate equilibrium shape. Both these should
result in the decrease of the moment of inertia, but the star's
rigidity resists the stresses until the star cracks when the
stresses reach a critical value.
Such a ``starquake'', an analogy of earthquake, rearrange the
stellar matter, and thus decreases abruptly the inertia momentum.
This global starquake with mass $\sim M_\odot$ could reproduce
large vela-like glitches, while the crust ($\sim 10^{-5} M_\odot$)
quake of neutron stars cannot.
In addition, the starquake may excite precession.


{\em Asteroseismology?}
It is well known that oscillation modes can provide much
information on stellar interior, the apotheosis of which is of
helioseismology which has largely confirmed the main elements of
the standard solar model.
We have then a chance to test the solid strange star model by
asteroseismology.
According to the differences of restoring forces, oscillation
modes can be divided into $p$ressure-mode (by pressure gradient),
$g$ravitational mode (by buoyancy in gravity), $r$otation-mode (by
Coriolis force in rotating stars), and $s$hear-mode (by shear
force).
The essence of a solid object is none-zero shear modulus, $\mu\neq
0$, which means that $p$- and $s$-modes can exist, but $g$- and
$r$-modes do not.

Fortunately variety of pulsation modes of fluid neutron stars have
been investigated previously (e.g., Andersso \& Comer 2001), the
results of which may provide an order-of-magnitude insight into
fluid strange stars.
It is found (Andersso \& Comer 2001) that the glitches of pulsars
can excite the modes large enough to be detected in the future
gravitational-wave detector (EURO with or without photon
shotnoise) if they are in a superfluid state.
However no gravitational-wave can be observed if glitching pulsars
are solid strange stars in EURO since, especially, no $r$-mode
instability can occur there.
It is not sure whether significant gravitational-waves can be
observed in EURO soon after a supernova if solid quark matter is
possible, since we lack exact knowledge of melting temperature
$T_{\rm m}$. The dynamics of supernova gravitational-wave should
(not) be changed significantly if $T_{\rm m}>10$ MeV ($T_{\rm
m}<1$ MeV).
It is true that a nascent hot strange star may be fluidity, and
its spin frequency is limited due to the gravitational radiation
caused by $r$-mode instability (Madsen 1998). However for cold
recycled millisecond pulsars, if being solid, the upper limit of
rotation frequency could be much larger than that given by Madsen
(2000).

Asteroseismology may validate the elastic properties of solid
strange stars.
Quasi-periodic oscillations with $\sim$kHz (kHz QPSs) are found in
``neutron'' stars, but not in black hole candidates (BHCs) (van
der Klis 2000).
This might be an unique signature of the center ``neutron'' stars.
In addition, a recent study with improved technique of data
processing shows that QPOs of ``neutron'' stars are quite
different from that of BHCs; the former could be originated by
stochastic oscillations, but the later by periodic modulations (Li
\& Chen 2003).
Could the kHz QPOs be caused by the global oscillations of a solid
strange star?

We propose an alternative mechanism for kHz QPOs, that torsional
oscillations, rather than radial ones, could excite effectively
the transverse Alfv\'en waves in magnetospheres and then affect
the accretion rates, while the global oscillations are excited by
accretion of varying rates with randomicity.
The solution of free oscillations of a uniform elastic sphere can
be found in Garland (1979). The torsional oscillation frequency is
$\nu=x(l)v_{\rm s}/(2\pi R)$, with the velocity of shear waves
$v_{\rm s}=(\mu^{\rm SS}/\rho)^{1/2}$ and $x(l)$ is a function of
degree $l$ of spherical harmonics (\{$x(1)=5.8, 9.1, 12.3, ...$\}
and \{$x(2)=2.5, 7.1, 10.5, ...$\} for $l=1,2$, respectively).
If kHz QPOs originates in this way, we can obtain the shear
modulus for solid strange stars,
\begin{equation}
\mu^{\rm SS}=4\times 10^{32} x_{10}^{-2}
R_6^2\rho_{15}\nu_3^2~~{\rm erg/cm^3},
\label{muSS}
\end{equation}
where $x=x/10$, $R_6=R/(10^6$cm),
$\rho_{15}=\rho/(10^{15}$g/cm$^3)$, and $\nu_3=\nu/(10^3$Hz). That
$\mu^{\rm SS}>>\mu^{\rm NS}$ is unsurprising since quark matter is
much denser than normal one.


\section{Discussions}

The ideal of solid quark matter is proposed, and three addressed
ways to test it are based on its various astrophysical
implications.
There may be other consequences of solid quark matter. (1) strong
magnetic field play a key role in pulsar life, but there is still
no consensus on its origin. It is worth noting that a
ferromagnetism-like domain structure may appear in a solid strange
star which could be a microscopic mechanism of pulsar magnetic
fields. Actually Yang \& Luo (1983) found that the quark-cluster
phase with parallel spins is energetically favored due to the
color magnetic interaction, and that the correspondent Curie
temperature is $\sim 10^2$ MeV. The ferromagnetism issue is also
noted recently by Tatsumi (2000), who suggested that the
Hartree-Fock state with the inclusion of spin polarization shows a
spontaneous magnetic instability at low densities through the same
mechanism as in electron gas.
(2) the calculation of the cooling behavior of strange stars
should include the energy release as quark matter solidifies.

%


{\em Acknowledgments}:
I thank Prof. Chongshou Gao for the discussion of the possibility
of quark matter in solid state. The helpful suggestions from an
anonymous referee are sincerely acknowledged. This work is
supported by NSFC (10273001) and the Special Funds for Major State
Basic Research Projects of China (G2000077602).


\end{document}